# Proton Acceleration in Underdense Plasma by Ultraintense Laguerre-Gaussian Laser Pulse

Xiaomei Zhang[*],　Baifei Shen[†],　Lingang Zhang,　Jiancai Xu,　Xiaofeng Wang,　Wenpeng Wang ,　Longqiong Yi,　and Yin Shi

*State Key Laboratory of High Field Laser Physics, Shanghai Institute of Optics and Fine Mechanics, Chinese Academy of Sciences, Shanghai 201800, China*

**Abstract:**

Three-dimensional particle-in-cell simulation is used to investigate the witness proton acceleration in underdense plasma with a short intense Laguerre-Gaussian (LG) laser pulse. Driven by the $LG_{10}$ laser pulse, a special bubble with an electron pillar on the axis is formed, in which protons can be well-confined by the generated transversal focusing field and accelerated by the longitudinal wakefield. The risk of scattering prior to acceleration with a Gaussian laser pulse in underdense plasma is avoided, and protons are accelerated stably to much higher energy. In simulation, a proton beam has been accelerated to 7 GeV from 1 GeV in underdense tritium plasma driven by a $2.14 \times 10^{22}$ W/cm$^2$ $LG_{10}$ laser pulse.

---
[*] Author to whom correspondence should be addressed. Electronic mail: zhxm@siom.ac.cn.
[†] Author to whom correspondence should be addressed. Electronic mail: bfshen@mail.shcnc.ac.cn.

1. **Introduction**

Proton acceleration by the intense circularly polarized (CP) laser pulse (often $I \sim 10^{22}$ W/cm$^2$) in radiation pressure acceleration (RPA) regime has recently become the focus of research [1-13]. Simulations and theoretical analyses have shown that high-quality proton beam with Giga-electron volts (GeV, $10^9$ eV) can be generated from the interaction between a short laser pulse and a thin foil [2, 4, 9-11, 14]. Considering its huge potential applications, some promising works are on progress. Increasing the energy of accelerated protons by RPA is difficult because of decreasing acceleration gradient after the accelerated proton velocity approaches the light speed. In addition, many undesirable effects, such as multidimensional instabilities [15-17], occur during acceleration with long interaction time, even if the laser pulse can be well controlled. Therefore, protons of energy beyond 5 GeV are rather hard to be obtained through the RPA mechanism.

On the other hand, tens of GeV or even TeV high-quality proton beam can be generated in the so called sequential radiation pressure and bubble acceleration regime [18-22] by using a $10^{23}$ W/cm$^2$ laser pulse with two dimensional (2D) particle in cell (PIC) simulations. The plasma channel can help to form the twin bubble structure, and thus proton energy can be over 10 GeV [22]. This promising approach can compete with the conventional accelerator in obtaining the ultra-high energy protons when the laser technology meets the requirement in the near future and the same encouraging results as the electron acceleration in bubble is expected for protons, so it is necessary to study the detailed process.

We know that, to get high energy protons, there are two key stages, namely, the trapping and the continual acceleration. The former can be reached by RPA like in the sequential radiation pressure and bubble acceleration regime. It has been found that a quasi-monoenergetic plasma bunch of high energy density can be obtained by irradiating a currently available short laser pulse irradiating on a small hemispheric shell target [23], which can also be used in the trapping process in the bubble. For obtaining ion energy higher than 5 GeV, the second stage, continuous acceleration, is more critical and challenging. In the regular bubble, electrons can be well-confined on the acceleration axis because of the transverse focusing field and continuously accelerated in the laser propagation direction. However for protons, it is an opposite situation that protons will be dispersed by the radial electric field and thus difficult to be further accelerated without additional compensation method. We note the recent work has reported special donut wakefields driven by a LG laser pulse for positron and electron acceleration[24]. For positron acceleration, the laser pulse should be short enough to avoid affecting the positron because the acceleration and the laser fields coexist at the same space. However, the proton mass is much larger, so maybe it is more appropriate to use such a donut wakefield to accelerate protons because of the limited effect of the laser field on them.

In this paper, the well-confined acceleration of externally injected protons in the wakefield driven by Laguerre-Gaussian (LG) laser pulse is studied by using three-dimensional (3D) PIC simulations. Trapping conditions and acceleration results are analyzed. It is found that, different from the regular bubble induced by a Gaussian laser pulse, a special bubble with an electron pillar on the axis will be formed when a LG laser pulse irradiates in underdense plasma. This structure provides an efficient and strong focusing force for protons in the transverse direction; thus, protons can be accelerated continuously for a long time. The 3D PIC simulations confirm that a proton beam with initial energy of 1 GeV has been accelerated to 7 GeV in the underdense plasma with a density of $2.4 \times 10^{20}$ /cm$^3$. This acceleration is stimulated using a $LG_{10}$ laser pulse with power similar to the Super-Gaussian (SG) laser pulse with an intensity $2.14 \times 10^{22}$ W/cm$^2$.

## 2. Simulation and Analysis

The proposed method is demonstrated with 3D PIC simulation code (VORPAL) [25]. The simulation box is 60 μm($x$)×100 μm($y$)×100 μm($z$), which corresponds to a moving window with $600 \times 300 \times 300$ cells and one particle per cell. The tritium underdense plasma occupies the $12 \, \mu m < x < 1000 \, \mu m$ region in the propagation direction of the laser pulse and $-35 \, \mu m < y(z) < 35 \, \mu m$ with a density of $n_0 = 2.4 \times 10^{20}$ /cm$^3$. The used CP $LG_{10}$ laser pulse is described as $a = a_0 \left( \sqrt{2} r / r_0 \right) \exp\left( -r^4 / r_0^4 \right) \exp(il\phi) \sin^2\left( \pi t / (2t_0) \right)$ with $r_0 = 7$ μm and $t_0 = 12.5T$, where $T$ is the laser period and $\phi$ is the azimuthal angle within the range of $[0 \, 2\pi]$. The laser electric field is used with power similar to that of the SG laser pulse with peak intensity of $2.14 \times 10^{22}$ W/cm$^2$, which corresponds to its normalized amplitude

$a_0 = eA/m_e c^2 = 70.7$ for the laser pulse with wavelength $\lambda = 0.8$ μm, where $A$ is the vector potential, $c$ is the light speed in vacuum, $m_e$ is the electron mass and $e$ is the electron charge. At $t = 0$, the laser pulse enters the simulation box from the left boundary. Here, the slow-moving massive background tritium ions allow the formation of a stable electron bubble with a large space-charge field [26]. The witness proton beam emitted from $x = 5$ μm, $y = z = 0$ μm is of the size of $3\mu m \times 1\mu m \times 1\mu m$ and with total charge of 160 pC.

The background electrons are expelled by an intense laser pulse through the pondermotive force, which is proportional to $\nabla I$. In a regular bubble driven by a Gaussian laser pulse, the transverse field will disperse protons located near the *x* axis and push them to the bubble sheath wall. However, in the case of LG laser pulse, the $LG_{10}$ laser pulse has a hollow-structure electric field as shown in Fig. 1, in which the field on axis is zero. Therefore, electrons expelled outward by the pondermotive force of the laser pulse in transversal direction form the outer bubble sheath, whereas the electrons expelled inward form the inner bubble sheath. That is, a bubble structure with an electron pillar on the *x* axis will be formed when it propagates in the underdense plasma because of its transverse donut-like shaped intensity[24, 27]. The electron pillar on the *x* axis of the special bubble structure will result in a focusing field for protons around the inner electron pillar. If the transverse laser ponderomotive force is sufficiently intense to compress the inner electron pillar to an electron thread with high density, the wakefield can trap and accelerate protons for a long time. In this case, this structure provides the focusing force for protons in transverse direction and the accelerating force in longitudinal direction. With the parameters in this paper, a clear special bubble structure with high density electron thread is formed, as shown in Fig. 2. The charge density of the electron thread is higher than the background ion density. Also there is a high density electron bunch trapped in the rear of the special bubble, which will not be described because it is beyond the scope of the present context.

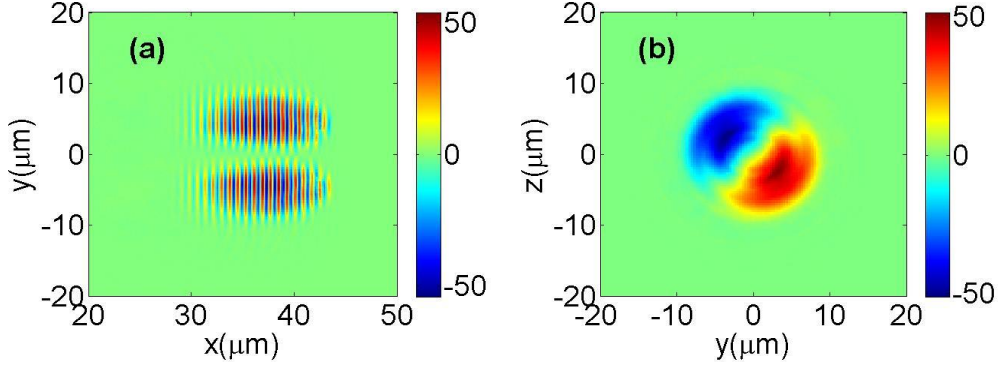

Fig. 1. Field distributions of the LG$_{10}$ laser pulse at $t$=160 fs are given (a) in the $x$-$y$ plane at $z$=0 and (b) in the $y$-$z$ plane at $x$=39 μm. The field is normalized to $m_e \omega_0 c / e$, where $\omega_0$ is the laser frequency.

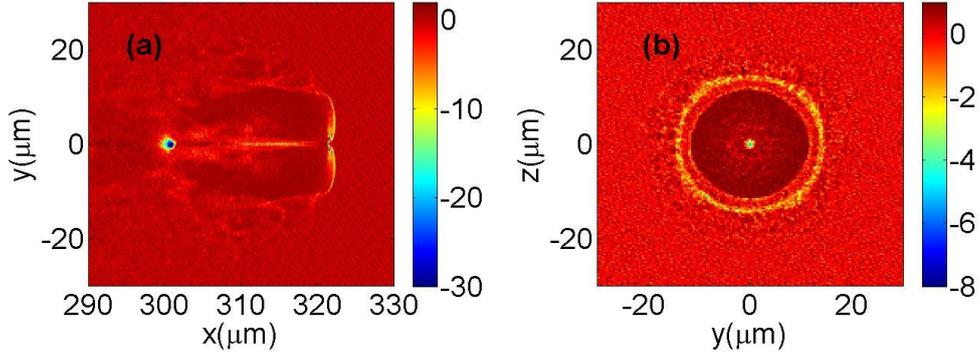

Fig. 2. Contours of the charge density at $t$=1.12 ps are given (a) in the $x$-$y$ plane at $z$=0 and (b) in the $y$-$z$ plane at $x$=314.2 μm. The charge density is normalized to $en_0$.

The electric fields at $t$= 1.12 ps in longitudinal and transverse directions in the special bubble are shown in Fig. 3 from the 2D simulation with constant physical parameters in order to observe the fields clearly. The longitudinal wakefield around the $x$ axis is about $6m_e\omega_0 c/e$, which can accelerate protons. More important is that there is indeed a transverse focusing field as shown in Fig. 3(b), which is expressed by $\sim (E_y - cB_z)$. Simultaneously, the witness proton beam is located in the black-dashed box and confined by the focusing field. The focusing field is similar to the twin bubbles in previous study [22], but with different generation schemes. In the present study, the structure is axial symmetry decided by the laser pulse mode and can propagate stably for a long distance for proton acceleration. As expected, protons are well-confined transversely on the axis in the special bubble driven by the LG$_{10}$ laser pulse, as shown in Fig. 4(a) with the momentum distribution at $t$=1.12 ps (most protons are located in a bucket within the radius of $r$=1μm in transverse direction), and accelerated stably to $t$=2.13 ps when they start dephasing (surpassing the bubble front). Eventually, the witness protons gain the peak energy of 7 GeV as shown in Fig. 4(b). In the case of a SG laser pulse, keeping other simulation parameters unchanged, the protons diffuse gradually and cannot be accelerated stably, Fig. 4(c) shows that the proton beam filling with the simulation box (out of the bubble). Finally, the peak energy stops at 2.2 GeV.

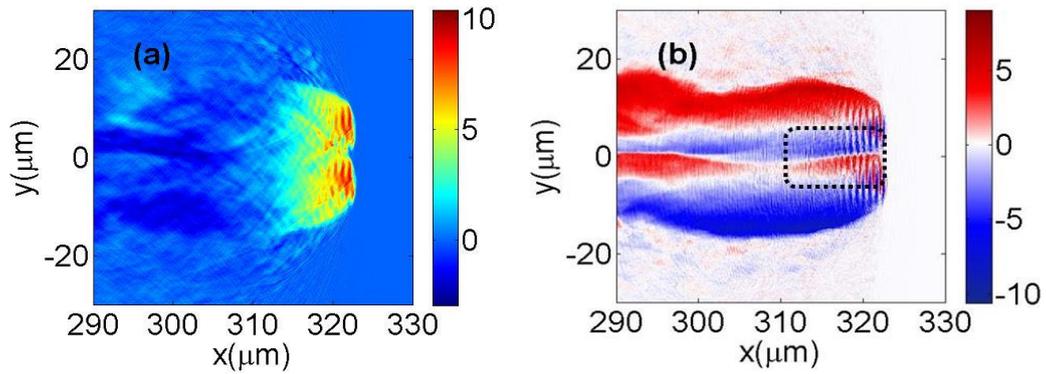

Fig. 3. (a) Acceleration field in the laser propagation direction, and (b) the focusing field in transverse direction are shown at $t$= 1.12 ps. The field in the black-dashed box has a focusing effect on protons in transverse direction. The field is normalized by $m_e\omega_0 c/e$.

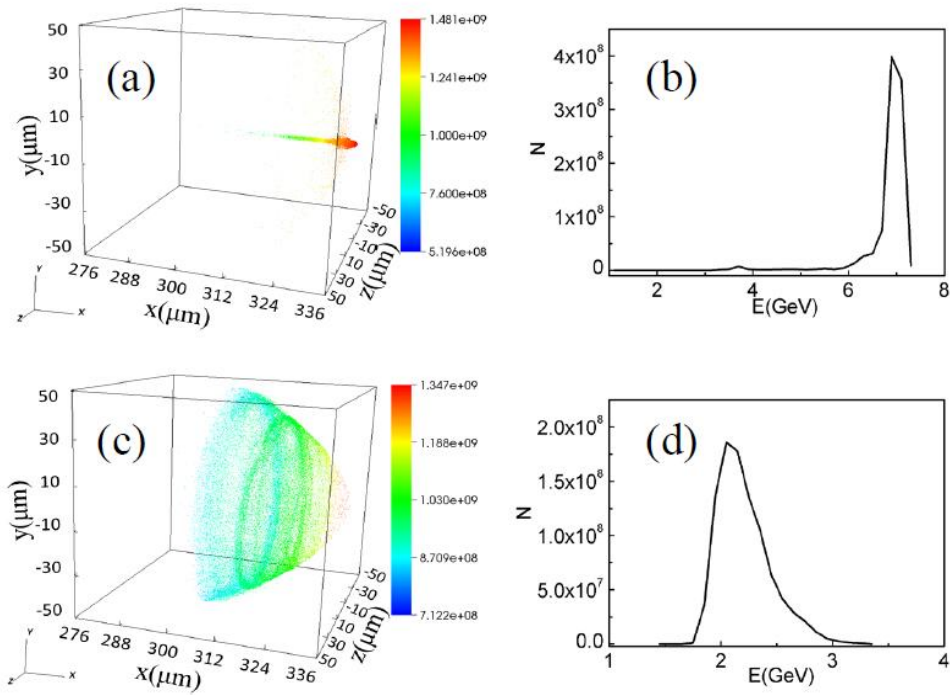

Fig 4. Longitudinal momentum distributions of protons at *t*=1.12 ps driven by (a) the LG$_{10}$ laser pulse and (c) SG laser pulse, and the energy spectra of the witness protons driven by (b) the LG$_{10}$ laser pulse at *t*=2.13 ps and (c) the SG laser pulse at *t*=1.12 ps. The color represents the momentum value in the unit of $\gamma v_x$, where $\gamma$ and $v_i$ are the relativistic factor and longitudinal velocity of protons, respectively. *N* is the real proton number in (b) and (d).

## 3. Discussion

There are two important aspects for the present acceleration approach. First, we should note the electron pillar in the special bubble structure, which determines the longitudinal acceleration field for the witness proton beam, is crucial for the acceleration. If the electron pillar is too thick, the wakefield around the witness protons will be too weak to accelerate the protons because the charge separation field is almost neutralized. Therefore, laser intensity, laser spot size, and plasma density should be adjusted to be appropriate to make the inner electron pillar thin enough to make the longitudinal acceleration field intense enough. Assuming the bubble is spherical-like, its radius (the distance between the bubble wall and the electron thread) can be estimated by balancing the transverse laser ponderomotive force on a single electron and the transverse charge separation field force [28]. The transverse ponderomotive force of the CP laser pulse is $F_{pond} = e^2 / (4 m_e \omega_0^2) \nabla E^2$, where $E = m_e \omega_0 c a / e$. The transverse charge separation field force is $E_{trans} = 4\pi e^2 n R$, where $R$ is the bubble radius. Then,

$$(R/\lambda) = (a_0 / 2\pi)\sqrt{n_c / 2n}, \qquad (1)$$

where $n_c = m_e \omega_0^2 / (4\pi e^2)$ is the critical plasma density. Fig. 2 shows that the radius $R \sim 15 \mu m$ is consistent with Eq. (1). The radius minimally changes because of the strong non-linear effects. On the other hand, for the ultraintense LG laser pulse, the special bubble radius is also related to the laser spot size for its shaped intensity and requires $r_0 \sim R/2$. This relationship is important for the formation of the special bubble which is appropriate to accelerate protons, because in this case, the inner electron pillar can be compressed to the thin electron thread and its thickness can be neglected. On the present of the thin electron thread, the focusing force

$F_t = e(E_y - cB_z) \sim enr_0$ in the transverse direction is induced

Another important aspect is the trapping and energy gain of witness protons. The initial energy of the witness protons required for trapping depends on the potential in the special bubble (intensity of the acceleration field). The Hamiltonian of a proton in the wakefield and electromagnetic field is $h_0 = \sqrt{1 + p_x^2 + \rho_p^2 a^2} - \rho_p \phi(\xi) - v_p p_x$, where $\xi = x - ct$ and $h_0$ is an integration constant corresponding to the initial proton condition in front of the bubble, $p_x = \gamma v_x$ is the proton longitudinal momentum, $\rho_p = -1/1836$, $\phi(\xi) = \int E_x d\xi$ is the scalar potential, $E_x$ is the acceleration field with peak intensity estimated roughly using $E_{x\max} \sim \sqrt{a_0}$ for the Gaussian pulse[28], and $v_p$ is the bubble velocity related to the plasma density and the laser intensity. For the ultraintense LG laser pulse, the wakefield on the axis is slightly weaker due to the transverse shaped intensity. For simplicity, we define the longitudinal wakefield increases linearly within $0 < \xi < R$ and deceases linearly within $R < \xi < R + d_{skin}$ with $E_{x\max} \sim \eta \sqrt{a_0}$, where the wakefield is zero at $\xi = 0$, and $d_{skin} \sim 1/\sqrt{n_e}$ is the skin width of the high density electron sheet in the front of the bubble, and $\eta \leq 1$ is a factor. Then, the motion of protons can be given by

$$p_x = \frac{v_p \left( \rho_p \phi(\xi) + h_0 \right) \pm \sqrt{\left( \rho_p \phi(\xi) + h_0 \right)^2 - \left( 1 - v_p^2 \right) \left( 1 - \left( \rho_p \phi(\xi) + h_0 \right)^2 \right)}}{1 - v_p^2}. \quad (2)$$

For the ultraintense CP laser pulse interacting with the underdense plasma, the high density electron sheet in front of the laser pulse expelled by the intense light pressure is overdense and the laser pulse is reflected. By balancing the momentum of the electron layer with the light pressure and considering the Doppler effect[29], $2I/c(1 - v_p/c)/(1 + v_p/c) = 2\gamma_p^2 n_e m_e v_p^2$, the bubble velocity can be obtained as follows:

$$v_p = \frac{a\sqrt{n_c/n_e}}{1 + a\sqrt{n_c/n_e}}, \quad (3)$$

where $n_e$ is the electron layer density which can be obtained by $n_e \sim n_0 R / d_{skin}$. From Eq. (3), $v_p = 0.96$ agrees well with that ($v_p \sim 0.967 \, (\gamma_p = 3.9)$) from the above simulation. The protons are accelerated from the onset of their injection into the bubble from the bubble front. Until the proton velocity becomes equal to the bubble velocity, the protons are trapped in the bubble and accelerated continually. The trapping condition can be written as follows:

$$h_0 + \rho_p \phi_{trap} = 1/\gamma_p, \tag{4}$$

where $\gamma_p = 1/\sqrt{1-v_p^2}$ and $\phi_{trap}$ is the potential required for trapping. For the protons injected with an initial positive velocity $v_0$, $h_0 = \gamma_0 (1 - v_p v_0)$ where $\gamma_0 = 1/\sqrt{1-v_0^2}$.

The potential required for trapping ($\phi_{trap}$) and the energy after the positive acceleration field ($\gamma_{end}$) with different initial energies ($\gamma_0$) of the witness proton are shown in Fig. 5(a), in which $\eta = 2/3$ according to the simulation results. With the increasing initial energy of the witness proton, the required potential decreases. The trapping of the protons depends on the maximum potential $\phi_{max} = \int_0^{R+d_{skin}} E_x dx$. With these simulation parameters, $\phi_{max} \simeq 380$, which corresponds to the initial energy $\gamma_{0min} \simeq 1.4$. For protons with initial energy lower than $\gamma_{0min}$, they will leave behind the acceleration phase and cannot be trapped, such as the protons at rest initially requiring the potential of $\phi_{trap} = 1368$ which is higher than $\phi_{max}$. These protons shown in red line in Fig. 5(a) are not fully accelerated, and their end energy is not high as expected. For protons with the initial energy higher than $\gamma_{0min}$, they will be trapped and accelerated in the wakefield, such as the protons with $\gamma_0 \simeq 2$ in the above simulation requiring the potential of $\phi_{trap} = 128$ which is smaller than $\phi_{max}$. These protons shown in blue line in Fig. 5(a) move with the bubble and gain high energy. The analysis result is in well agreement with the simulation results. Of course, the higher initial energy, the shorter acceleration time before they are dephased and the less energy they gain. Therefore, the maximum potential of the bubble play a key role for the trapping and acceleration. Fig. 5(b) shows that the maximum potential ($\phi_{max}$) changes with the laser amplitude ($a_0$) for different underdense plasma densities. That is to say, in order to obtain the high $\phi_{max}$, we may choose high laser intensity and low plasma density. One point should be noted in this case that the bubble velocity will be higher and so the required potential for trapping will be higher from Eqs. (3) and (4).

The analysis shows that the LG laser pulse generates the focusing force in transverse direction, which is crucial for the proton acceleration in the wakefield. Although the acceleration field around the *x* axis is slightly weakened, the protons with an appropriate initial energy which is related to the maximum potential of the acceleration field still can be trapped and accelerated stably, and get nearly ten orders energy gain.

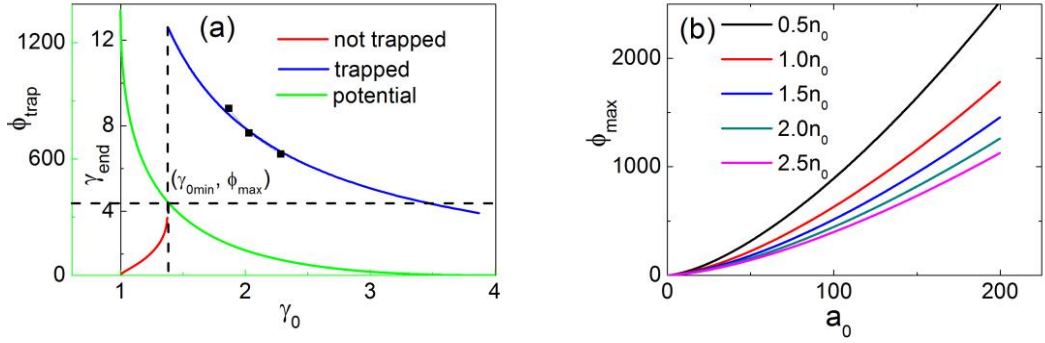

Fig 5. (a) Distributions of the potential required for trapping ($\phi_{trap}$) and the energy after accelerated ($\gamma_{end}$) with different initial energies ($\gamma_0$) of the witness proton. The black squares indicate the simulation results, which is in well agreement with the analysis results. (b) The changes of the maximum potential ($\phi_{max}$) with the laser amplitude ($a_0$) for different underdense plasma densities.

## 4. Conclusion

In conclusion, the witness proton acceleration in the wakefield driven by a $LG_{10}$ laser pulse has been studied. Confining the protons near the acceleration axis is important for the continuous acceleration in bubble regime. By using the $LG_{10}$ laser pulse, a special bubble with a high density electron thread on the axis has been found, in which an enough intense acceleration field in longitudinal direction and a focusing field in transverse direction for the protons coexist. The 3D

PIC simulation results show that protons can be well-confined near the axis and accelerated stably for a long time. The energy of protons is increased to 7 GeV from 1 GeV.

One point should be noted that even though the initial radius of the witness proton bunch is much larger than that of the on-axis electron sheath, the proton bunch will be focused to the axis because of the existence of the transverse focusing field and the simulation results verifies it (not presented in this paper). Additionally, obtainable high energy protons with the currently available laser level is realistic because it is easier to realize, and our study show that the present scheme still works for the lower intensity, such as $a_0$ around 20, but requires witness protons with higher initial energy. Moreover, generation of such intense $LG_{10}$ laser pulse is critical for the present scheme. Fortunately, a potential approach has been proposed recently [30].

**ACKNOWLEDGMENTS**

This work has been supported by the Ministry of Science and Technology (2011CB808104 973, 2011DFA11300), the National Natural Science Foundation of China (Projects No. 11125526, No. 11335013, No. 11374319, No. 11374317, No. 11305236, No. 11127901, and No. 61221064), and the Shanghai Natural Science Foundation (No. 13ZR1463300).